\documentclass[12pt,a4paper,dvips]{article}
\usepackage{a4p}
\usepackage{cite,mcite}
\usepackage{graphicx}
\usepackage{physics }
\usepackage{l3_title,ifthen,Lep}
\journalname{Phys. Lett. B}
\date{October 1, 1999}
\preprint{99-138}
\Lep{2}
\newlength{\capindent}
\newlength{\capwidth}
\newlength{\figwidth}
\newcommand{\icaption}[2][!*!,!]{\hspace*{\capindent}%
  \begin{minipage}{\capwidth}
    \ifthenelse{\equal{#1}{!*!,!}}%
      {\caption{#2}}%
      {\caption[#1]{#2}}
  \end{minipage}}
\begin{document}
\begin{titlepage}
\title{Search for Excited Leptons at \boldmath{$\sqrt{s}=189\GeV$} }

\author{The L3 Collaboration}

%
%
\begin{abstract}
A search for excited leptons, 
$\rm e^*,~\mu^*, ~\tau^*, ~\nu_e^*, ~\nu_{\mu}^*$ and $\nu_{\tau}^*$,
was performed with the L3 detector at LEP using data collected at a
centre--of--mass energy of 189 \GeV, corresponding to an integrated 
luminosity of $\rm 176~pb^{-1}$. 
No evidence of their production is observed.
From the searches for pair produced excited leptons, lower mass limits 
are set close to the kinematic limit. 
From the searches for singly produced excited leptons, upper limits on 
their couplings are derived in the mass range up to 189 \GeV{}. 
\end{abstract}

\submitted
\end{titlepage}

%
%
\section{Introduction}

The existence of excited leptons would provide evidence for fermion
substructure. Composite models could explain the number of families and
make the fermion masses and weak mixing angles calculable\cite{yellow}.
The excited leptons $\rm e^*,~\mu^*, ~\tau^*$ (collectively denoted as
$\ell^*$) and $\nu^*$ have been 
extensively searched for at the LEP \ee{} and 
the HERA $\rm ep$ colliders\cite{LEP_HERA}.
These particles are assumed to have spin and isospin values 
$1/2$ and the same electroweak SU(2) and U(1)
gauge couplings to the vector bosons as the standard leptons,
but are expected to constitute both left and right handed weak 
isodoublets. Excited leptons are
expected to decay into their ground states
by radiating a photon or a massive vector boson.

At \ee~colliders, excited leptons can be produced either in pairs 
(\ee $\longrightarrow \ell^*\ell^*, \longrightarrow \nu^*\nu^*$) 
or singly (\ee $\longrightarrow \ell\ell^*, \longrightarrow \nu\nu^*$). 
In pair production, the coupling of the excited leptons to the gauge bosons
is described by the Lagrangian\cite{hagi}
$${\cal L}_{\ell^*\ell^*} = \bar{L^*} \gamma^{\mu} \left( g  
{\vec\tau \over 2} \vec W_{\mu} + {g'} {Y \over 2} B_{\mu} \right) L^*. $$
The cross section for pair production depends on the mass of the excited lepton
~\cite{yellow,neus}.

Single production as well as magnetic decay can be described by
means of an effective Lagrangian of the form\cite{hagi}
$${\cal L}_{\ell^*\ell} = 
{gf \over \Lambda}   {\bar L^*} \sigma^{\mu\nu} {\vec\tau \over 2} 
l_{L} \partial_{\mu}{\vec W_{\nu}} +
{g'f' \over \Lambda} {\bar L^*} \sigma^{\mu\nu} Y
l_{L} \partial_{\mu}B_{\nu} + h.c.,$$
where $\Lambda$ is the compositeness scale and $f$ and $f'$ are the couplings
associated with the SU(2) and U(1) gauge groups of the Standard Model, 
respectively. 
They determine the production rate
of single excited leptons and their branching ratio 
into standard leptons plus gauge bosons\cite{neus}. 
Table~\ref{tab:bratio} 
shows the decay branching ratios for excited leptons  for
two relative values of $f$ and $f'$ and for different excited lepton masses.

Excited leptons are searched for in the radiative decays,
$\rm \ell^* \rightarrow \ell \gamma$, $\rm \nu^* \rightarrow \nu \gamma$, 
and weak decays,
$\rm \ell^* \rightarrow \nu W$ and $\rm \nu^* \rightarrow \ell W$. In 
the production of pairs of excited leptons, the search is performed considering
only radiative decays or only weak decays, but not one excited lepton decaying 
radiatively and the other decaying weakly. Pair production searches are 
sensitive to excited leptons
of mass up to values close to the kinematic limit, {\it i.e.} the beam energy.
Single production searches extend the sensitivity 
to the mass range above the beam energy up to the centre--of--mass energy,
$\sqrt{s}$. 

%
%
\section{Data Sample and Event Simulation}

The data sample analysed corresponds to $\rm 176.4 ~pb^{-1}$
collected with the L3 detector\cite{l3-detector} at LEP
at $\sqrt{s}=188.6\GeV$ in 1998.
For the simulation of background from 
Standard Model processes, different Monte Carlo programs are used:
Radiative Bhabha events are generated using BHWIDE\cite{bhwide} and 
TEEGG\cite{tee}. For other radiative dilepton events, $\mu\mu\gamma$, 
$\tau\tau\gamma$ and $\nu\nu\gamma$, the KORALZ\cite{koralz} generator is
used. The GGG\cite{ggg} Monte Carlo is used for final states with only 
photons. KORALW\cite{koralw} is used for the \ee$\rm \rightarrow WW$ 
process. PYTHIA\cite{pythia} is used for $\rm qq(\gamma)$, $\rm ZZ$
and $\rm Zee$ production and EXCALIBUR\cite{exca} is used for the 
$\rm qqe\nu$ final state.

The generation of excited leptons and their decay is done 
according to their differential cross section~\cite{hagi},
to optimise the selections and estimate the efficiencies. 
Initial state radiation is not implemented in the
generation, but it is taken into account 
in cross section calculations.
The generated events are passed through the L3 detector 
simulation\cite{geant}, which includes the effects of energy loss, 
multiple scattering, interactions and decays in the detector 
and the beam pipe.

%
%
\section{Radiative Decays}

Excited leptons decaying radiatively give rise to final states with low 
multiplicity and high energy photons. Event selection criteria reject
pure hadronic events, keeping a high signal efficiency 
independent of the flavour of excited leptons. This is achieved by accepting
events with less than eight tracks and at least one photon with energy greater
than $15\GeV$ in the central detector region 
($|\cos{\theta_\gamma}| < 0.75$, where 
$\theta_\gamma$ is the photon polar angle). To
reject cosmic background at least one scintillator within $\pm 5$ ns 
of the beam
crossing must be present. In addition, events with muons are required to have
a muon track pointing to the primary vertex.

Electromagnetic clusters are identified as electrons if there is a
matching track within $5^\circ$ in the $r\phi$ projection.
Muons are identified from tracks in the muon chambers.
A minimum ionising particle in the calorimeters can be accepted as a 
second muon.
The tau identification is based on jets constructed from calorimetric 
clusters, tracks and muons, with invariant mass below $3 \GeV$ and at least 
one associated track.
Events with charged leptons and photons in the final state are subjected to a
kinematic fit which imposes energy and momentum conservation. This improves
the resolution in the invariant mass of the $\rm \ell \gamma$ pairs.
These selection criteria are complemented by
additional requirements specific to the production channel and
the excited lepton flavour.
%
%
\subsection{Pair Production}

In order to select candidates for  
\ee$\rm\longrightarrow \ell^*\ell^* \longrightarrow \ell\ell\gamma\gamma$, it 
is further required that two photons with
energy greater than $15\GeV$, at least one of them in the central
region, and two lepton candidates of
the same flavour must be present. The difference of the two lepton-photon 
masses is required to be
smaller than $10\GeV$ and their sum greater than $100\GeV$.

Event selection for
\ee$\rm\longrightarrow \nu^*\nu^* \longrightarrow \nu\nu\gamma\gamma$ is 
based on a signature with only two energetic photons in the final state.
The preselection is complemented by 
requiring  two photons in the central region
with energies in the range
$0.2 < {\textstyle E_{\gamma}}/{\textstyle E_{beam}} < 0.8$
and with the acoplanarity angle greater than $10^\circ$.
There should be no tracks in the central tracker or in the muon chambers
and the energy of calorimetric clusters other than photons must
be smaller than $5\GeV$.

The number of observed events, the expected background and the signal 
selection efficiency are reported in Table~\ref{tab:selections}.
The main background is due to radiative dilepton events
\ee$\rm \rightarrow \ell\ell\gamma\gamma, \nu\nu\gamma\gamma$. 

%
%
\subsection{Single production}

Event selection for
\ee$\rm\longrightarrow \ell\ell^* \longrightarrow \ell\ell\gamma$ identifies
final states with two leptons and one photon with
energy greater than $20\GeV$.
At least one of the two possible $\rm \ell \gamma$ invariant
masses must be greater than $70\GeV$.
Events with just an identified electron and a photon, with invariant mass
above $70\GeV$, are also accepted for the excited electron selection. Thus,
a high signal efficiency is kept for the signal events originating from 
the $t$-channel exchange where one electron escapes along the beam pipe. 

Final states from 
\ee$\rm\longrightarrow \nu\nu^* \longrightarrow \nu\nu\gamma$ are 
characterised by a single photon with 
energy greater than $0.15 \sqrt{s}$. 
Neither tracks in the tracking chamber, nor in the muon chambers should
be present in the event. To reject cosmic events, no more than eight
calorimetric clusters must be present, and besides the photon, none 
of them should exceed $5 \GeV$.

The number of observed events, the expected background and signal 
selection efficiency are reported in Table~\ref{tab:selections}.
The main background is due to radiative dilepton events
\ee$\rm \rightarrow \ell\ell\gamma, \nu\nu\gamma$. 

%
%
\section{Weak Decays}

Weakly decaying excited leptons, $\rm \ell^* \rightarrow \nu W$ and 
$\rm \nu^* \rightarrow \ell W$, with at least one $\rm W$ decaying 
hadronically give rise to high multiplicity final states. Selection
criteria are designed to reject leptonic and two-photon events.
More than three tracks and 14 calorimetric clusters are required in
events with visible energy above $60\GeV$.

The lepton identification is the same as for low multiplicity events
except taus, which are identified as low multiplicity charged jets
satisfying at least two of the following four conditions: less than four
tracks associated with a jet, less than five calorimetric
clusters, invariant mass smaller than $2\GeV$,
or at least 70\% of its energy contained in a  $5^\circ$
half-opening cone.
Final selection of the candidates requires
additional criteria specific to the production channel and
the excited lepton flavour.

%
%
\subsection{Pair Production}

Final states from
\ee$\rm\rightarrow \ell^*\ell^* \rightarrow \nu\nu W W$ and
\ee$\rm\rightarrow \nu^*\nu^* \rightarrow \ell\ell W W$ 
are similar to Standard Model WW production. 
Due to the large cross section, pair production of excited 
leptons would manifest itself as an enhancement of the measured WW 
cross section. A combination of four selections, denoted by 
$\rm qqqq$, $\rm qqe$, $\rm qq\mu$ and $\rm qq\tau$, is used. To 
achieve a high signal efficiency, no attempt is made to reject 
the WW background. 

For the $\rm qqqq$ selection, events with at least three charged jets 
and visible energy greater than $140\GeV$ are selected.
For at least one pair of jets,
their invariant mass must be in the range between $50\GeV$ and 110 GeV,
the recoiling
mass against these two jets must be greater than $50\GeV$ and the sum of 
invariant plus recoil masses must be greater than $120\GeV$.
To reduce the $\rm q q \gamma$ background, 
events with missing momentum above 
$30\GeV$ at low polar angle, $|\cos{\theta_{miss}}|>0.95$, where 
$\theta_{miss}$ is the polar angle of the missing momentum, or
an electromagnetic cluster with energy above $50\GeV$ are rejected.

For the $\rm qq\ell$ selections, events with missing momentum greater 
than $10\GeV$, at high polar angle $|\cos{\theta_{miss}}|<0.95$, 
and the difference between visible energy and missing momentum
smaller than $165\GeV$ are selected. 
In addition an isolated lepton with energy greater than $5\GeV$ is required.
The invariant mass of the event (without the identified lepton) must be
in the range between $40\GeV$ and 120 \GeV.

In the search for excited electron and muon neutrinos, the signal sensitivity
is enhanced by requiring two additional isolated electrons or muons in the
event, with energy in the range between $3\GeV$ and 15 \GeV. 
Table~\ref{tab:selections} reports the yields of the selections.
The background is mainly due to WW and $\rm qq(\gamma)$ production.

%
%
\subsection{Single Production}

Experimental signatures of
\ee$\rm\rightarrow \ell \ell^* \rightarrow \ell \nu  W$ and
\ee$\rm\rightarrow \nu  \nu^*  \rightarrow \nu  \ell W$ 
are also similar to Standard Model WW production. The 
hadronic decays of the W are considered, and the $\rm qq\ell$
selections described above are applied.

They are complemented with a $\rm qq$ selection 
which requires two acoplanar hadronic jets with invariant mass 
in the range between $60\GeV$ and $100\GeV$ and recoil mass below $70\GeV$, 
to cope with undetected leptons lost at low angle or releasing low energy.

Events are selected as candidates due to single excited leptons if they 
pass any of the $\rm qq\ell$ or $\rm qq$ selections.
The main backgrounds are due to $\rm WW$, $\rm qqe\nu$  and 
$\rm qq (\gamma)$ production.
The combined efficiency for $\rm \ell^*$ and $\nu^*$ 
depends slightly on the mass and flavour of the
excited lepton, except in the case of the excited electron, 
in which the efficiency increases from 
9\% to 47\% in the $\rm e^*$ mass range between $95\GeV$ and 185 \GeV, as
shown in Table~\ref{tab:selections}.

%
%
\section{Results and Limits}

\begin{figure}[htb]
  \begin{center}
    \includegraphics[width=0.49\textwidth]{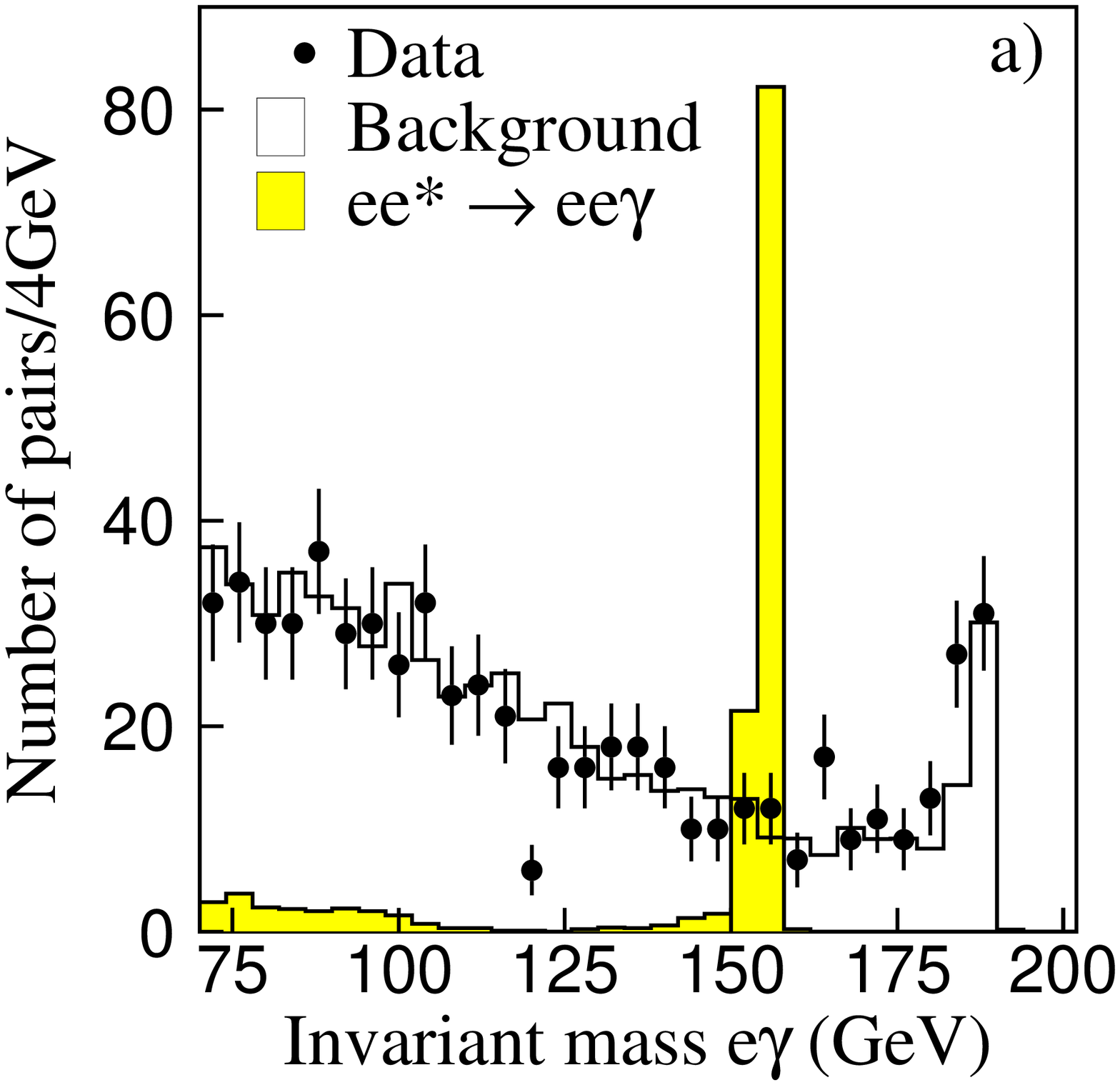}
    \includegraphics[width=0.49\textwidth]{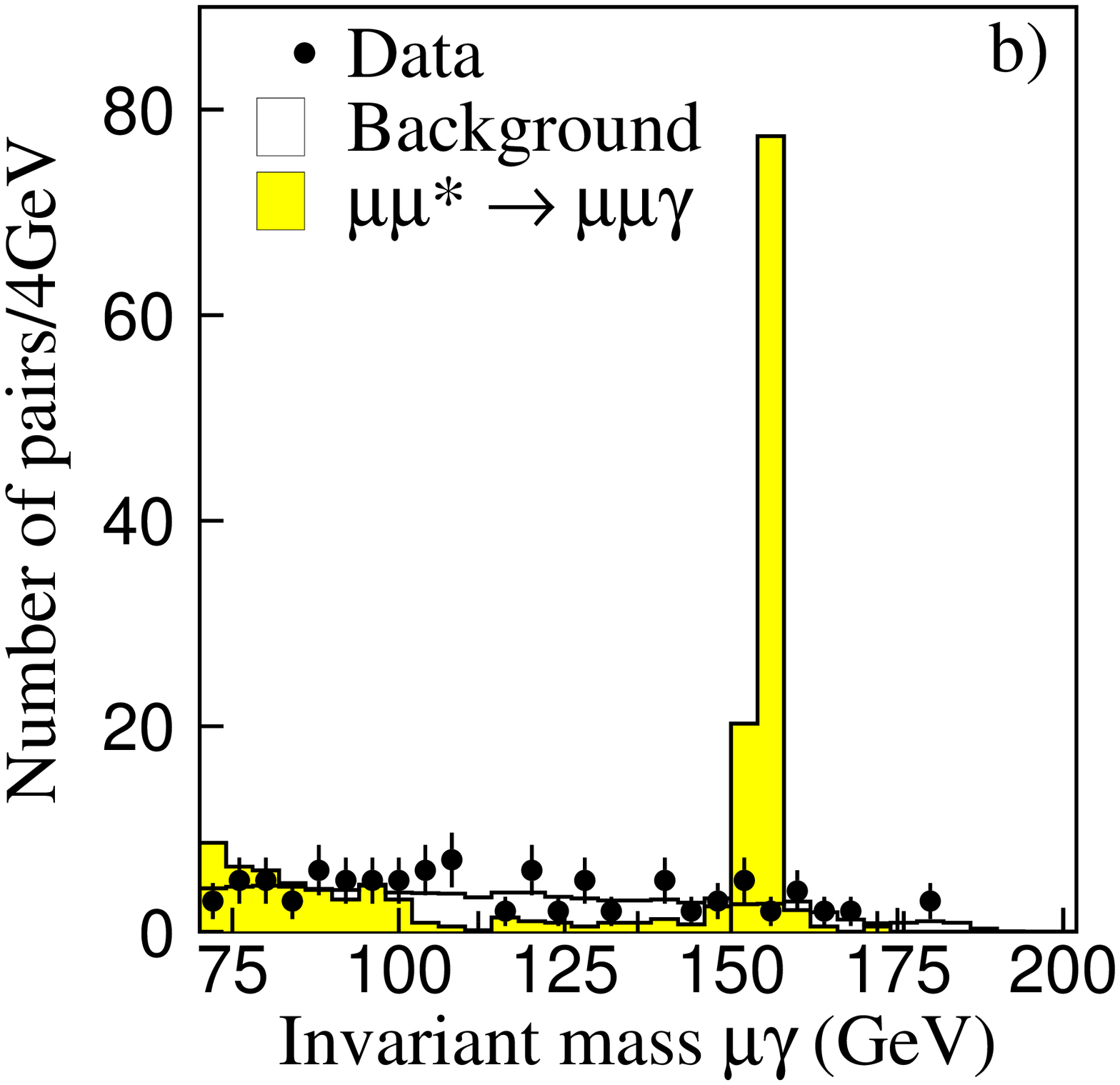}
    \includegraphics[width=0.49\textwidth]{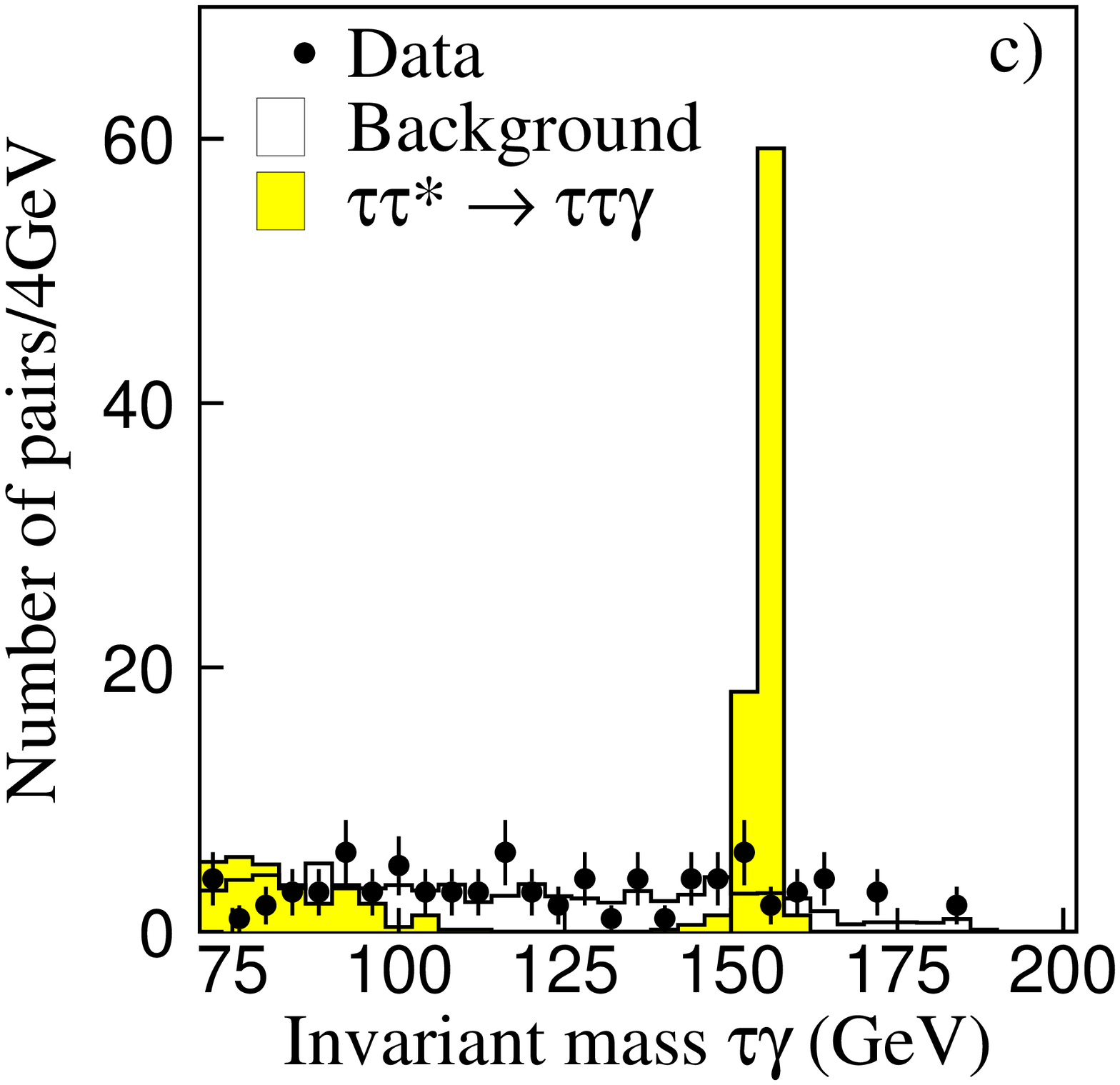}
    \includegraphics[width=0.49\textwidth]{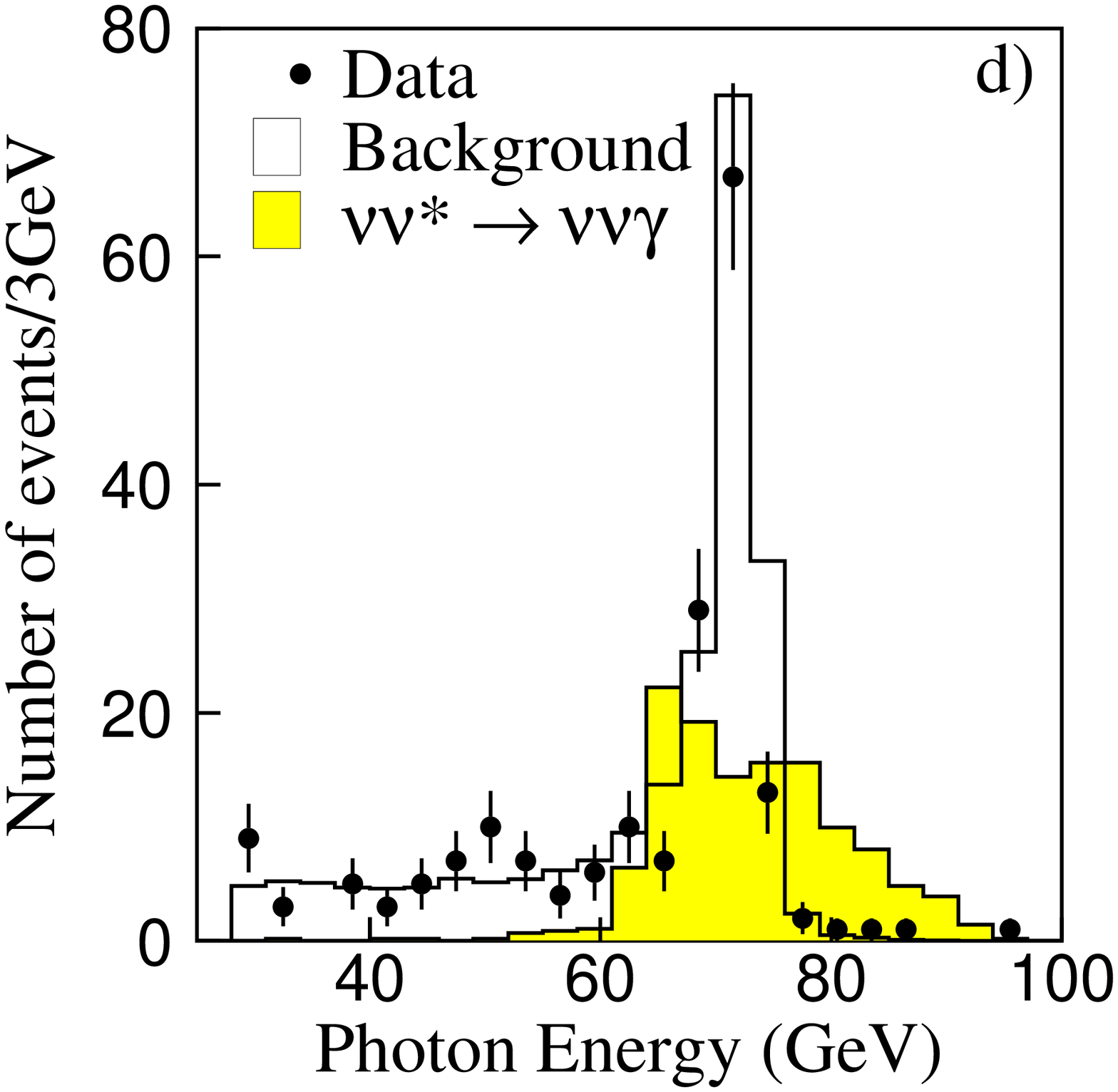}
  \end{center}
  \icaption{The invariant mass distributions for (a) e$\gamma$,
           (b) $\mu\gamma$, and (c) $\tau\gamma$ pairs. Energy distribution
           of single photon events (d).
           The expected signal
           for an excited lepton with a mass of $155 \GeV$  
           normalised to 1.0 pb for all the channels is shown together with data and Standard
           Model background.
  \label{fig:rad}}
\end{figure}

\begin{figure}[htb]
  \begin{center}
    \includegraphics[width=0.40\textwidth]{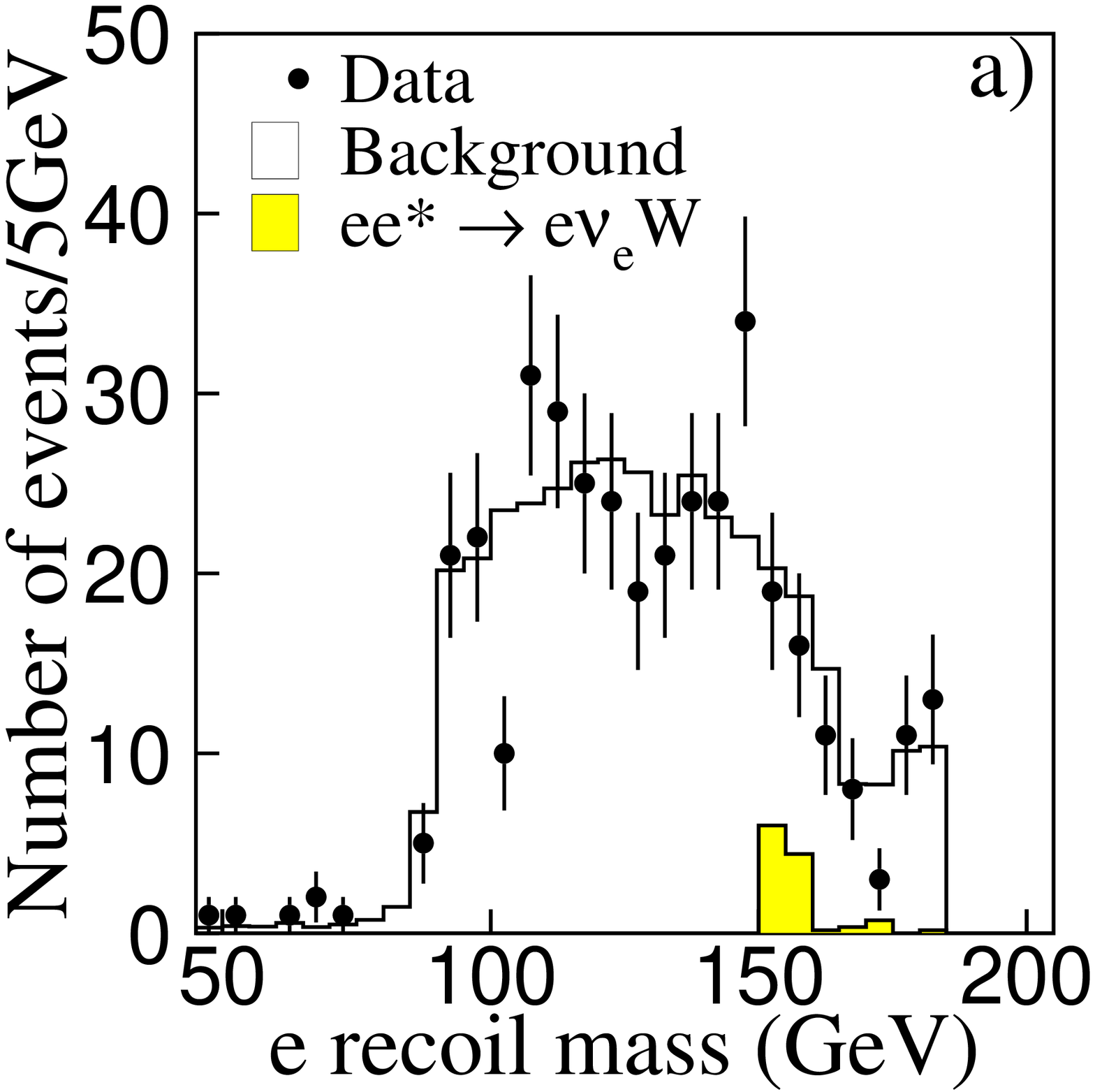}
    \includegraphics[width=0.40\textwidth]{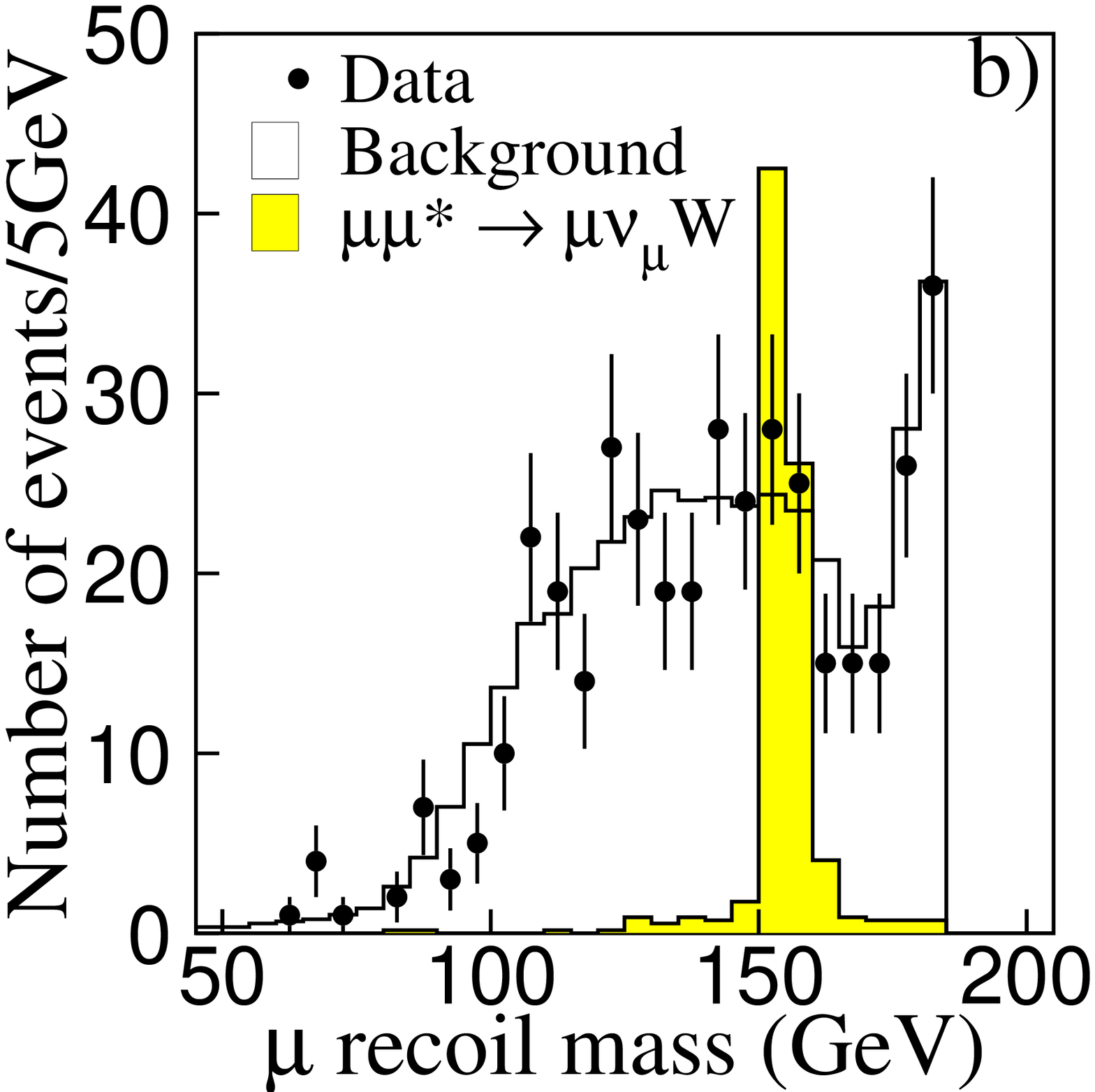}
    \includegraphics[width=0.40\textwidth]{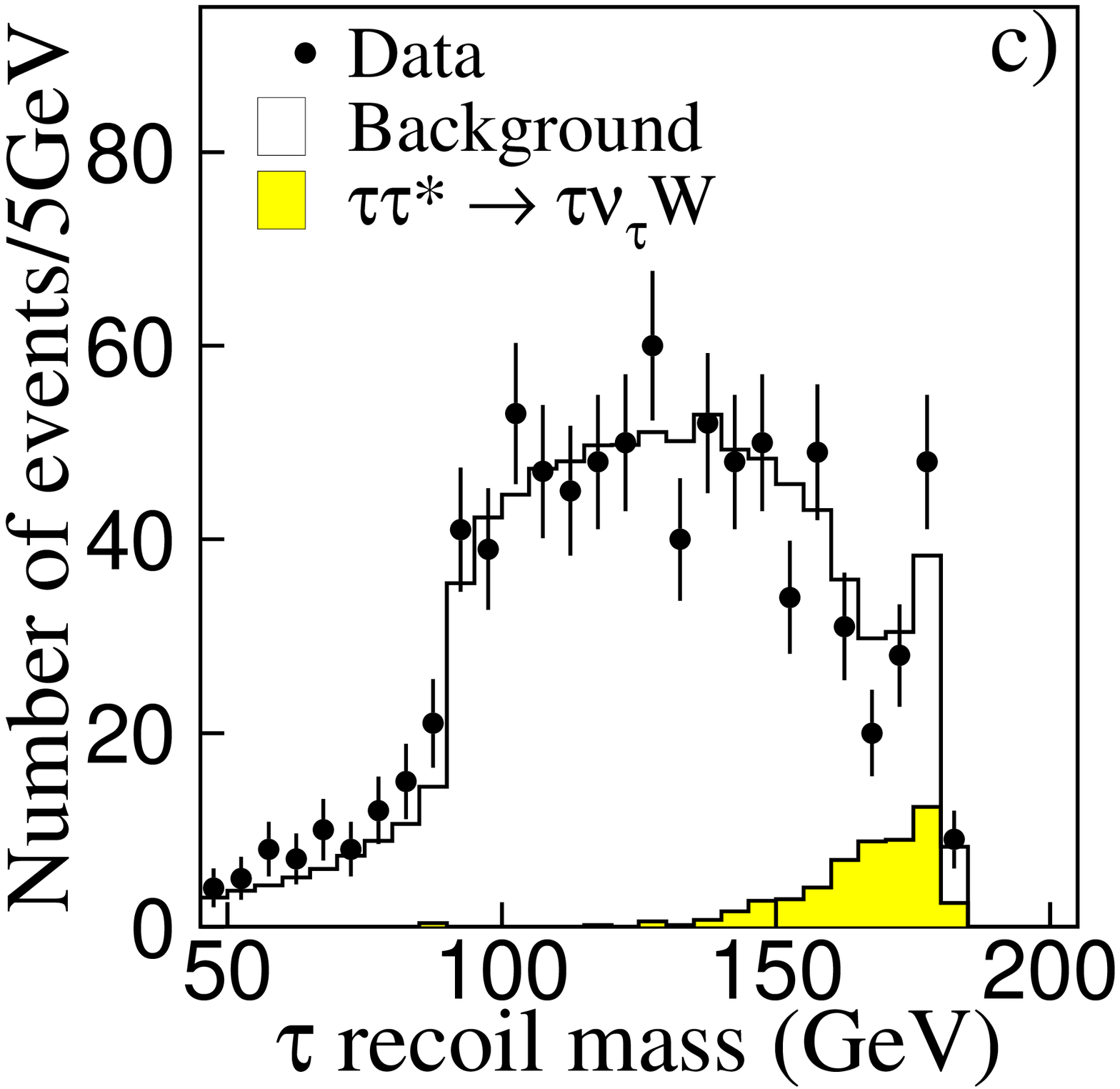}
    \includegraphics[width=0.40\textwidth]{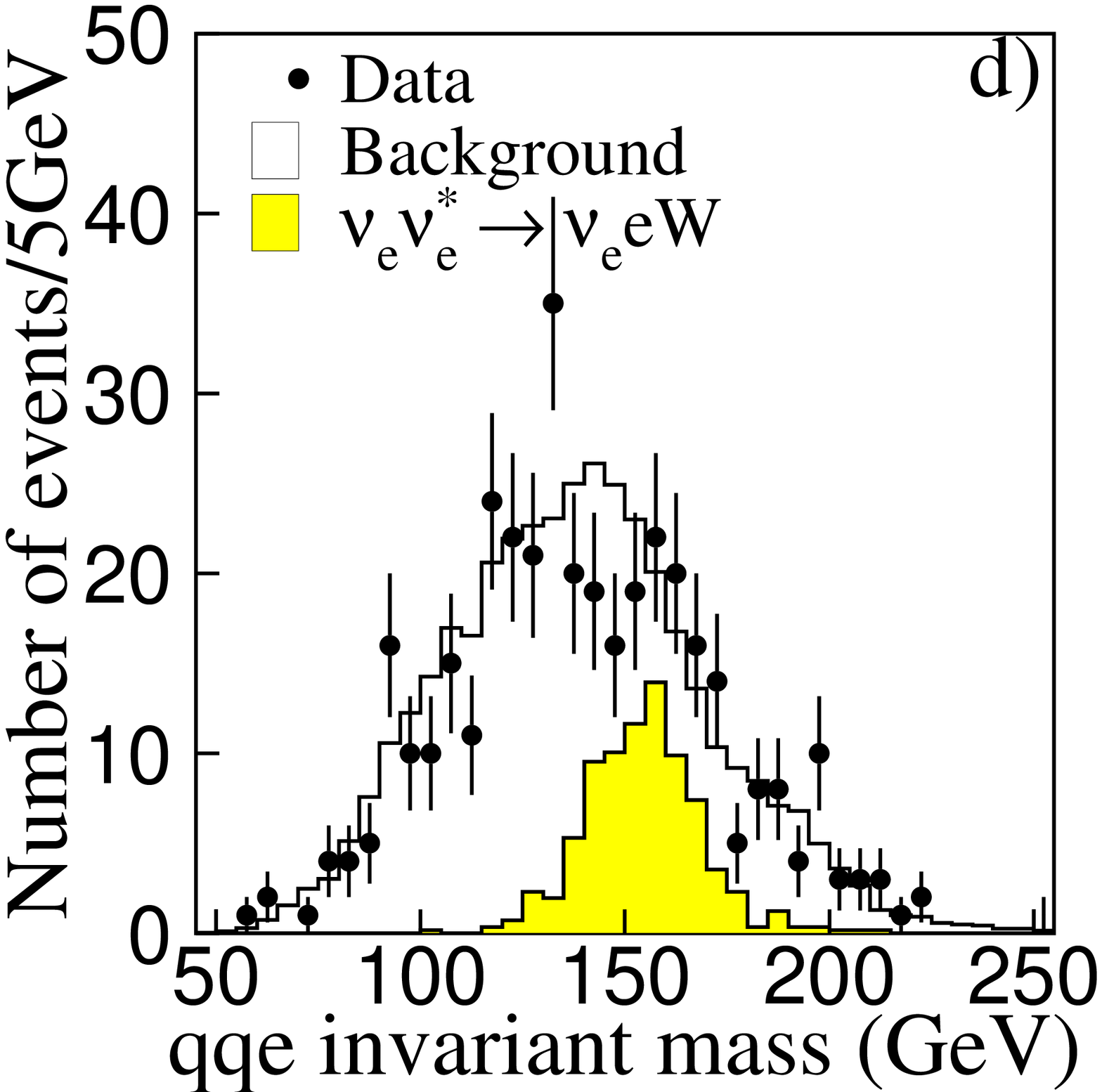}
    \includegraphics[width=0.40\textwidth]{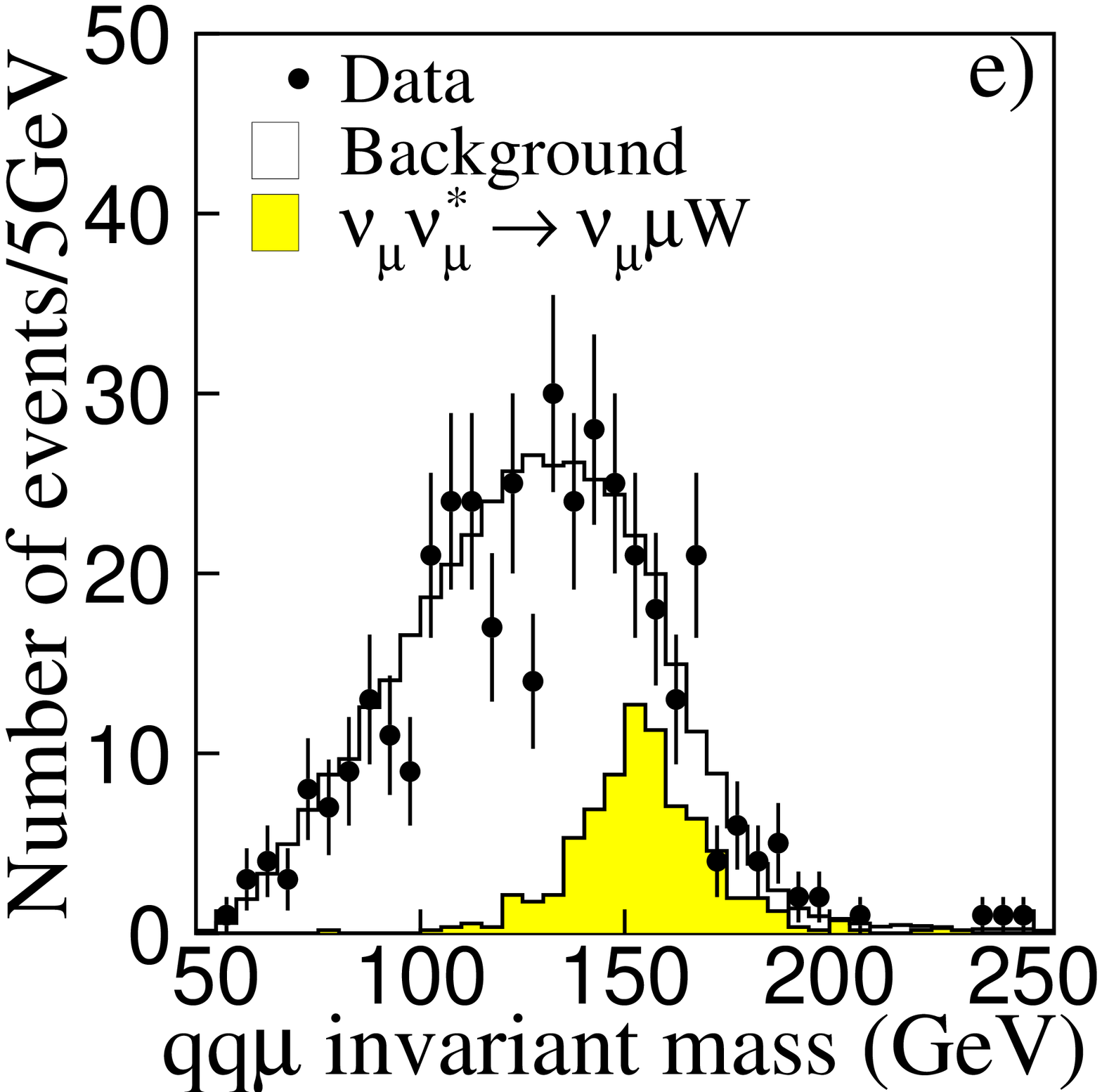}
    \includegraphics[width=0.40\textwidth]{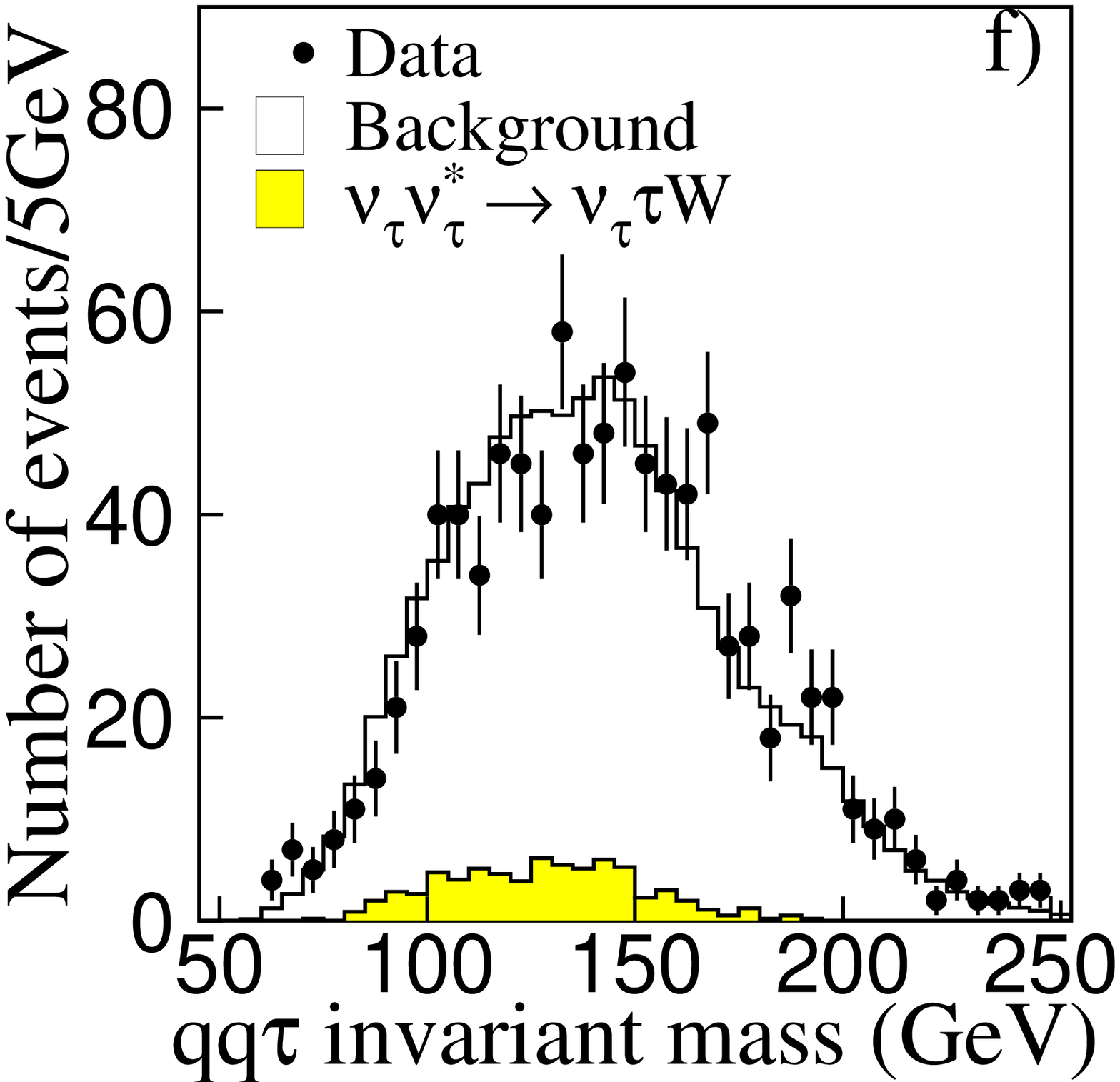}
  \end{center}
  \icaption{Recoil mass distributions for (a) electron , (b) muon , 
           and (c) tau, in the $\rm qq\ell$ selections. 
           Invariant mass distributions for (d) $\rm qqe$, (e) $\rm qq\mu$, 
           and (f) $\rm qq\tau$, selected events.
           The expected signal
           for an excited lepton with a mass of $155 \GeV$ 
           normalised to 1.0 pb for all the channels is shown together with data and Standard
           Model background.
  \label{fig:weak}}
\end{figure}

Figures~\ref{fig:rad}a-c show all 
combinations of invariant masses
$m_{{\rm e}\gamma}$, $m_{\mu\gamma}$ and $m_{\tau\gamma}$  above 
$70\GeV$.
Figure~\ref{fig:rad}d shows the 
energy of the photon in the single photon selection. 
Figures~\ref{fig:weak}a-c show 
the recoil mass against the identified lepton in the $\rm qqe$, 
$\rm qq\mu$ and $\rm qq\tau$ selections; {\it i.e.} the mass of the $\ell^*$
candidate.
Figures~\ref{fig:weak}d-f show the event invariant mass for the $\rm qqe$, 
$\rm qq\mu$ and $\rm qq\tau$ selections; {\it i.e.} the mass of the $\nu^*$
candidate. 
The number of observed events is consistent with the expected Standard Model
background, see Table~\ref{tab:selections}.
For each flavour and mass of excited lepton hypothesis, 
the number of candidates found in data
and expected from background are calculated. Taking into 
account the luminosity, the branching ratio 
(which depends on the ratio of the couplings $f / f'$) 
and the efficiency, an upper limit to the signal cross section is derived.
This limit is set at 95\% confidence level, using 
Bayesian statistics and assuming a flat {\it a priori} distribution for the signal.
In the case of pair production, the cross section only depends
on the mass and charge of the excited lepton. 
A lower mass limit is derived from the cross section limit.
In the case of single production, the cross section depends 
on the flavour and mass of the excited lepton and the value of the couplings 
$f$ and $f'$. The cross section limit is then interpreted in terms of an
upper limit to the coupling constant as a function of the mass of
the excited lepton.

Two different scenarios are
considered in order to calculate limits. In the first one, $f=f'$,
the radiative decay is dominant for
charged excited leptons whereas it is forbidden for excited neutrinos.
In the second one, $f=-f'$, the radiative decay is forbidden for
charged excited leptons whereas it is dominant for excited neutrinos.
Mass limits, as well as upper limits to the coupling constant, 
are thus derived from radiative decays for charged excited leptons
in the first case and excited neutrinos in the second case, and from
weak decays in the other searches.

The results from pair production searches in both the radiative and the
weak decay searches are combined to derive lower limits on the 
mass of excited leptons independent of the coupling constants $f$ and
$f'$. A scan is performed for all the possible ratios of the two 
couplings: ${\textstyle f}/{\textstyle f'} = \tan{\theta}$
with $\theta$ in the range 0 to $\pi$. 
For each value of  ${\textstyle f}/{\textstyle f'}$, 
the corresponding decay fractions 
$\rm \ell^* \ell^* \rightarrow \ell \ell \gamma \gamma$ and
$\rm \ell^* \ell^* \rightarrow \nu \nu W W$ 
($\nu^* \nu^* \rightarrow \nu \nu \gamma \gamma$ and
$\rm \nu^* \nu^* \rightarrow \ell \ell W W$ in the case of excited neutrinos)
are calculated, and a mass limit is set.  In Table~\ref{tab:limits} the 
lowest limit for each flavour of excited lepton is quoted. 

\begin{figure}[htb]
  \begin{center}
    \includegraphics[width=0.49\textwidth]{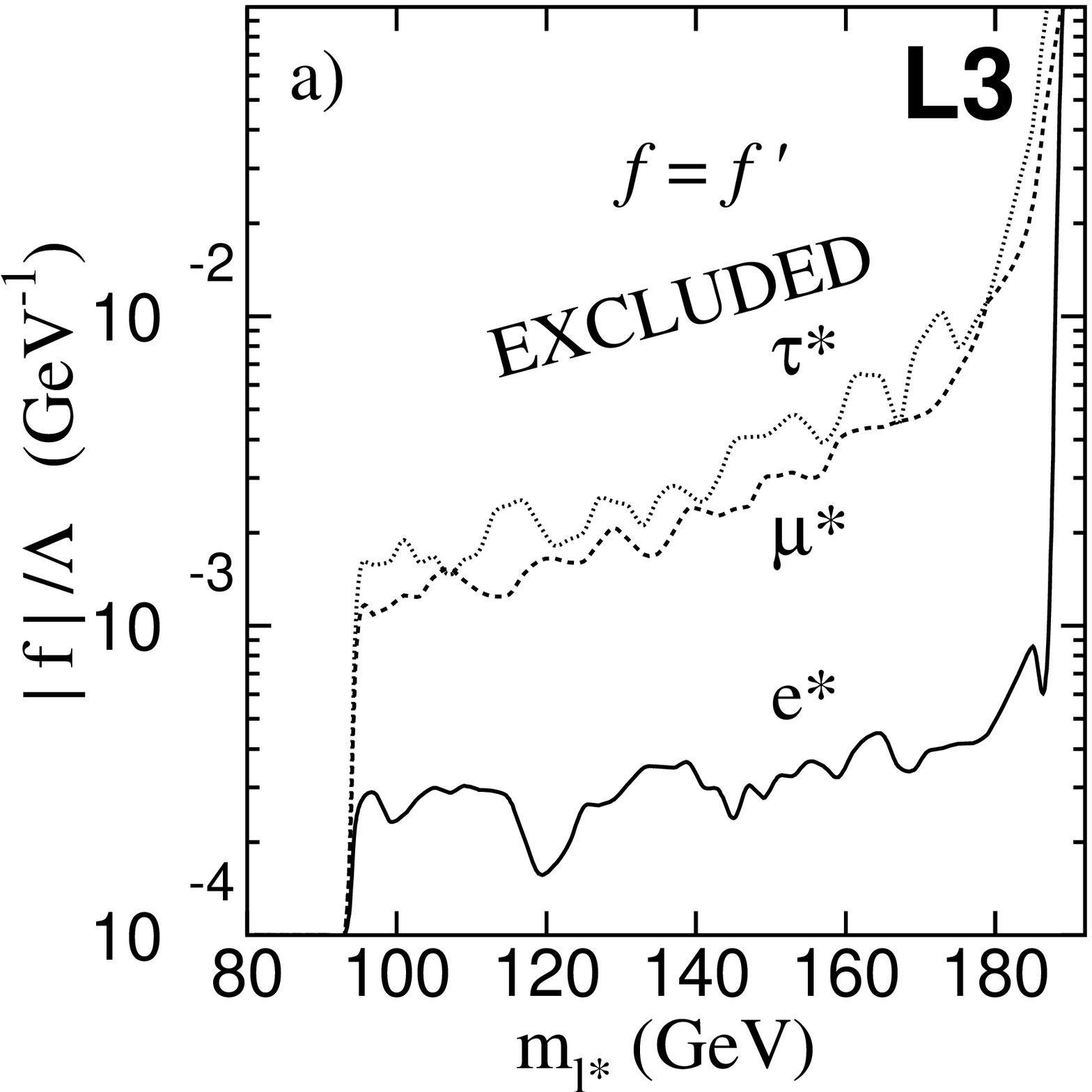}
    \includegraphics[width=0.49\textwidth]{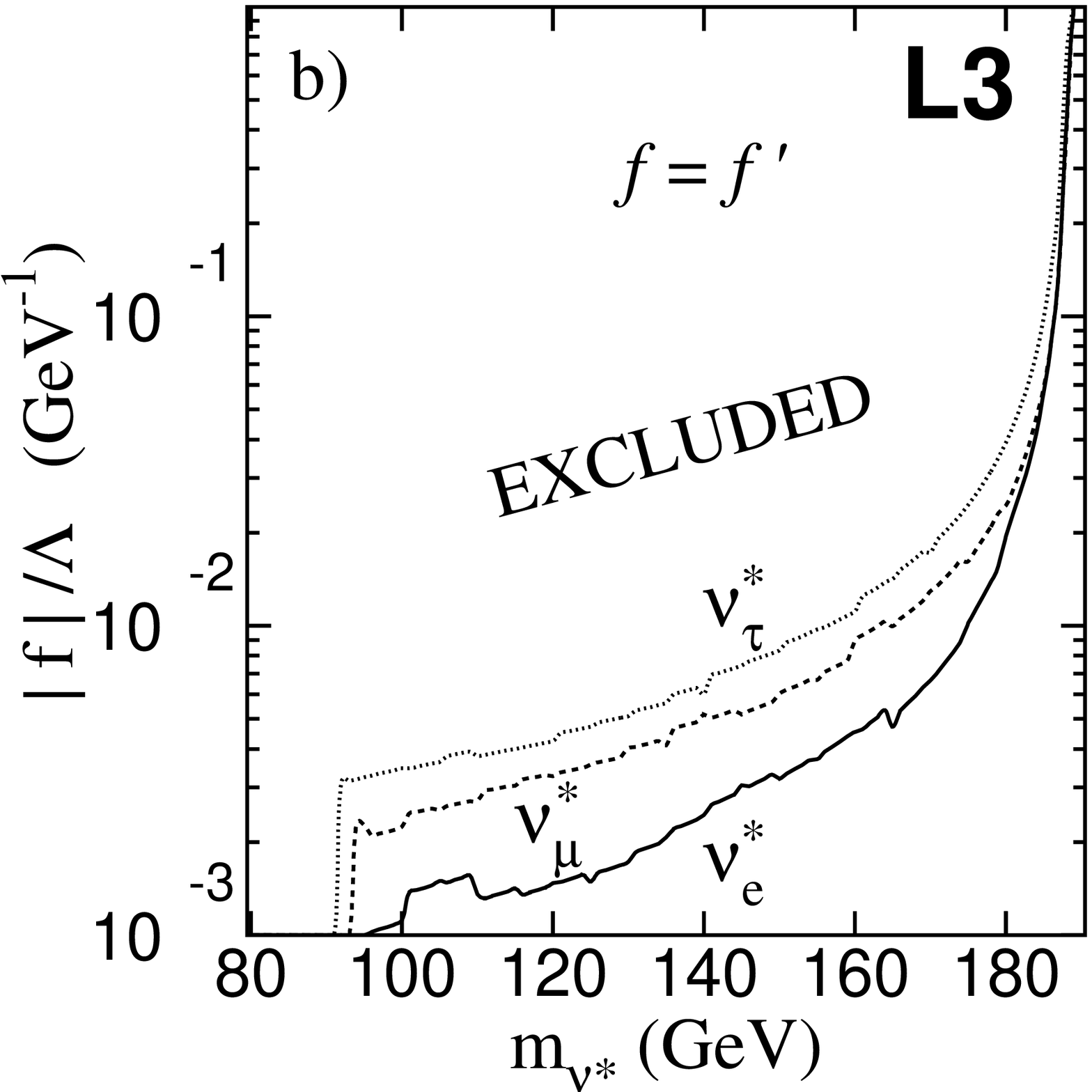}
    \includegraphics[width=0.49\textwidth]{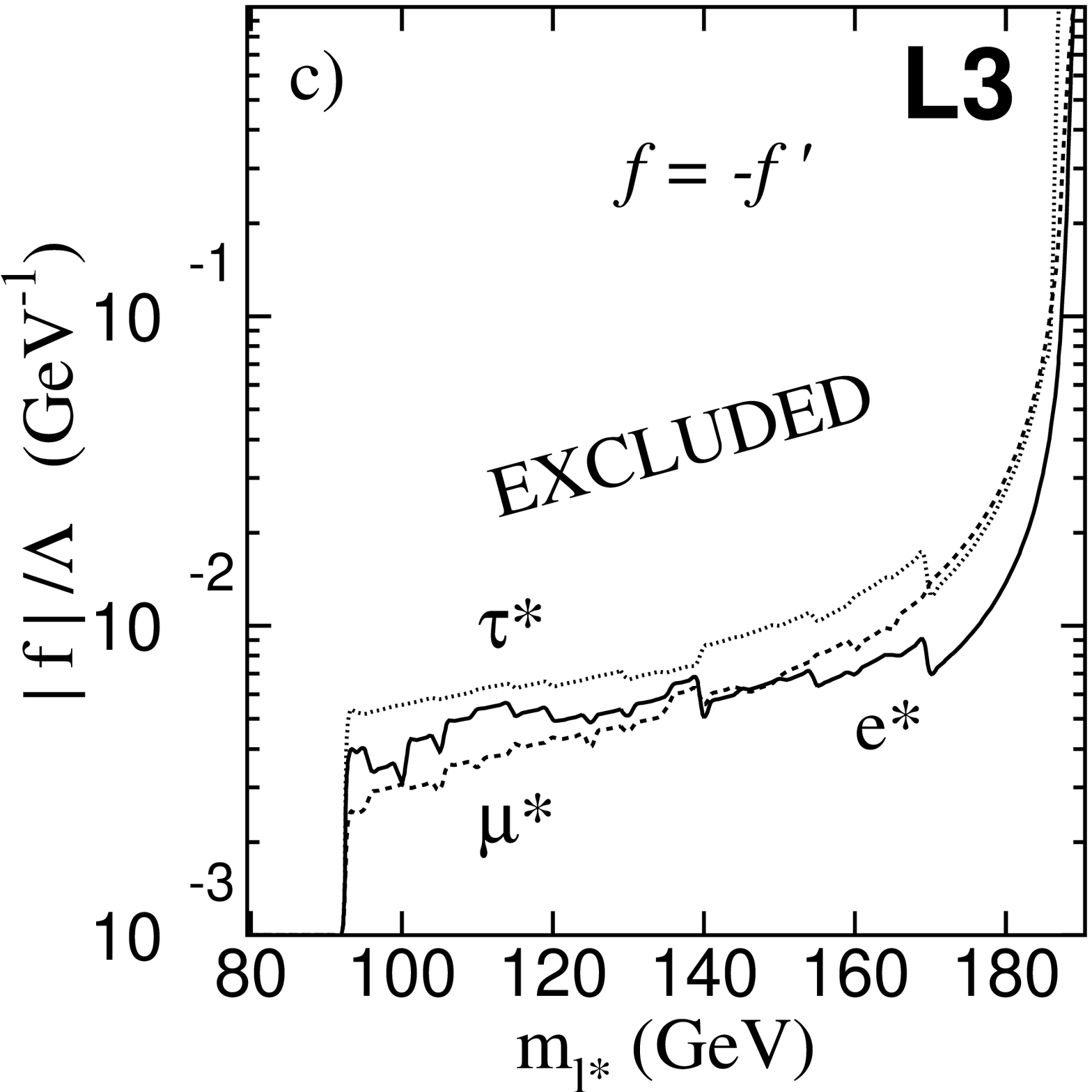}
    \includegraphics[width=0.49\textwidth]{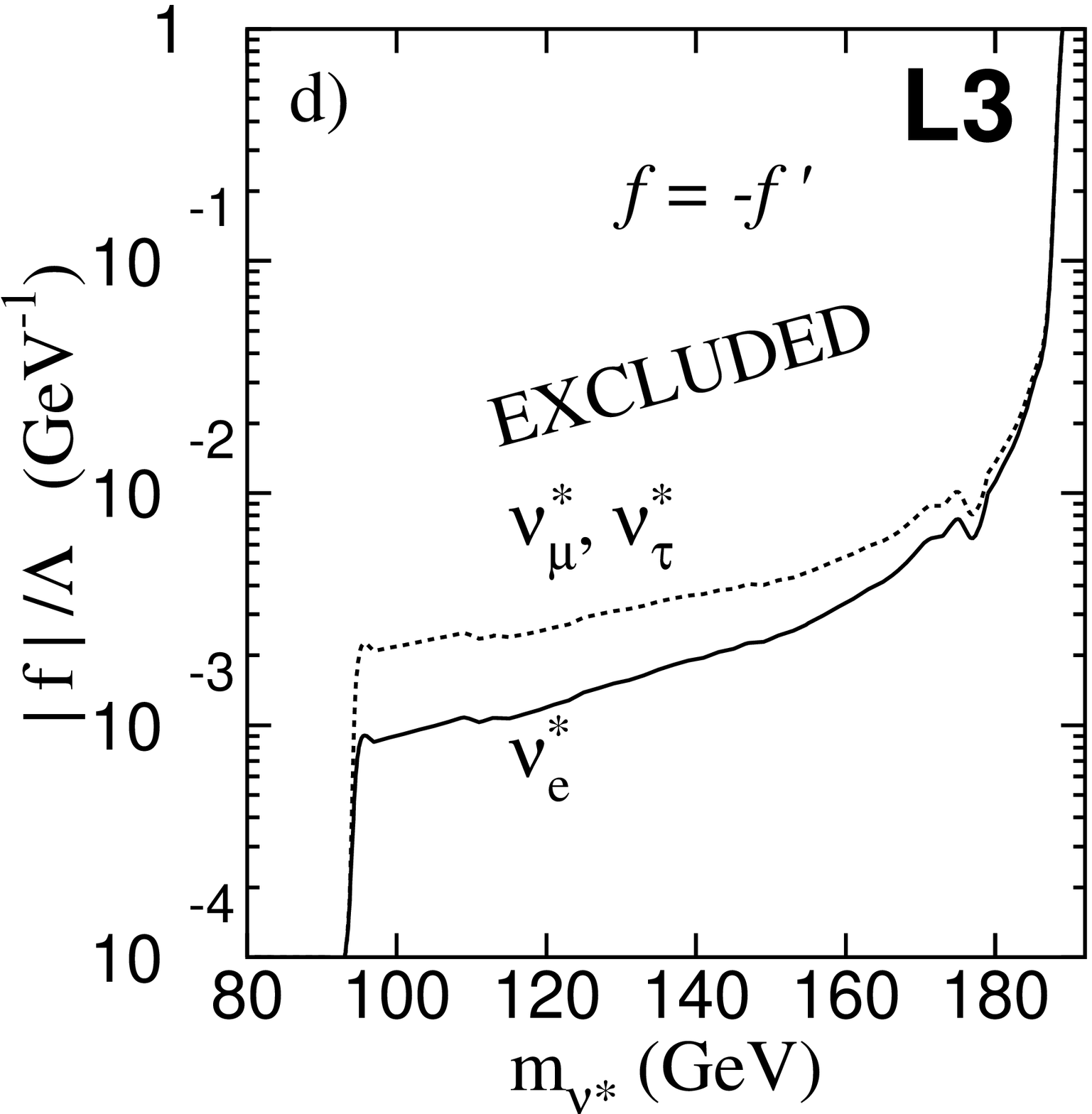}
  \end{center}
  \icaption{95\% confidence level upper limit on the coupling constant 
           ${\textstyle |f|} / {\textstyle \Lambda}$,
           as a function of the excited lepton mass  with $f=f'$:
      (a) $\rm e^*$, $\mu^*$ and $\tau^*$ ,
      (b) $\rm \nu_e^*$, $\nu_{\mu}^*$ and $\nu_{\tau}^*$,
           and with $f=-f'$:
      (c) $\rm e^*$, $\mu^*$ and $\tau^*$,
      (d) $\rm \nu_e^*$, $\nu_{\mu}^*$ and $\nu_{\tau}^*$.
  \label{fig:limites}}
\end{figure}

Figure~\ref{fig:limites} shows the upper limits to the
coupling constant for charged excited leptons and excited neutrinos
in both scenarios. The left-hand edge
of the curves indicates the lower mass limit derived from 
pair production searches, which is quoted in Table~\ref{tab:limits}. An
absolute lower mass limit of the order of 90 \GeV{} and upper limits
on the couplings in the mass range from 90 \GeV{} to 180 \GeV{} are set.
%
%
\section*{Acknowledgments}

We wish to express our gratitude to the CERN accelerator divisions for the excellent
performance of the LEP machine. We acknowledge the effort of the engineers
and technicians who have participated in the construction and maintenance of the 
experiment.

%
%

%
%
\newpage
\typeout{   }     
\typeout{Using author list for paper 191 -?}
\typeout{$Modified: Wed Sep 29 17:00:36 1999 by clare $}
\typeout{!!!!  This should only be used with document option a4p!!!!}
\typeout{   }
%
%
%
%
%
%

\newcount\tutecount  \tutecount=0
\def\tutenum#1{\global\advance\tutecount by 1 \xdef#1{\the\tutecount}}
\def\tute#1{$^{#1}$}
\tutenum\aachen            
\tutenum\nikhef            
\tutenum\mich              
\tutenum\lapp              
\tutenum\basel             
\tutenum\lsu               
\tutenum\beijing           
\tutenum\berlin            
\tutenum\bologna           
\tutenum\tata              
\tutenum\ne                
\tutenum\bucharest         
\tutenum\budapest          
\tutenum\mit               
\tutenum\debrecen          
\tutenum\florence          
\tutenum\cern              
\tutenum\wl                
\tutenum\geneva            
\tutenum\hefei             
\tutenum\seft              
\tutenum\lausanne          
\tutenum\lecce             
\tutenum\lyon              
\tutenum\madrid            
\tutenum\milan             
\tutenum\moscow            
\tutenum\naples            
\tutenum\cyprus            
\tutenum\nymegen           
\tutenum\caltech           
\tutenum\perugia           
\tutenum\cmu               
\tutenum\prince            
\tutenum\rome              
\tutenum\peters            
\tutenum\salerno           
\tutenum\ucsd              
\tutenum\santiago          
\tutenum\sofia             
\tutenum\korea             
\tutenum\alabama           
\tutenum\utrecht           
\tutenum\purdue            
\tutenum\psinst            
\tutenum\zeuthen           
\tutenum\eth               
\tutenum\hamburg           
\tutenum\taiwan            
\tutenum\tsinghua          
{
\parskip=0pt
\noindent
{\bf The L3 Collaboration:}
\ifx\selectfont\undefined
 \baselineskip=10.8pt
 \baselineskip\baselinestretch\baselineskip
 \normalbaselineskip\baselineskip
 \ixpt
\else
 \fontsize{9}{10.8pt}\selectfont
\fi
\medskip
\tolerance=10000
\hbadness=5000
\raggedright
\hsize=162truemm\hoffset=0mm
\def\r{\rlap,}
\noindent

M.Acciarri\r\tute\milan\
P.Achard\r\tute\geneva\ 
O.Adriani\r\tute{\florence}\ 
M.Aguilar-Benitez\r\tute\madrid\ 
J.Alcaraz\r\tute\madrid\ 
G.Alemanni\r\tute\lausanne\
J.Allaby\r\tute\cern\
A.Aloisio\r\tute\naples\ 
M.G.Alviggi\r\tute\naples\
G.Ambrosi\r\tute\geneva\
% ...
\newpage
\begin{list}{A}{\itemsep=0pt plus 0pt minus 0pt\parsep=0pt plus 0pt minus 0pt
                \topsep=0pt plus 0pt minus 0pt}
\item[\aachen]
 I. Physikalisches Institut, RWTH, D-52056 Aachen, FRG$^{\S}$\\
 III. Physikalisches Institut, RWTH, D-52056 Aachen, FRG$^{\S}$
\item[\nikhef] National Institute for High Energy Physics, NIKHEF, 
     and University of Amsterdam, NL-1009 DB Amsterdam, The Netherlands
\item[\mich] University of Michigan, Ann Arbor, MI 48109, USA
\item[\lapp] Laboratoire d'Annecy-le-Vieux de Physique des Particules, 
     LAPP,IN2P3-CNRS, BP 110, F-74941 Annecy-le-Vieux CEDEX, France
\item[\basel] Institute of Physics, University of Basel, CH-4056 Basel,
     Switzerland
\item[\lsu] Louisiana State University, Baton Rouge, LA 70803, USA
\item[\beijing] Institute of High Energy Physics, IHEP, 
  100039 Beijing, China$^{\triangle}$ 
\item[\berlin] Humboldt University, D-10099 Berlin, FRG$^{\S}$
\item[\bologna] University of Bologna and INFN-Sezione di Bologna, 
     I-40126 Bologna, Italy
\item[\tata] Tata Institute of Fundamental Research, Bombay 400 005, India
\item[\ne] Northeastern University, Boston, MA 02115, USA
\item[\bucharest] Institute of Atomic Physics and University of Bucharest,
     R-76900 Bucharest, Romania
\item[\budapest] Central Research Institute for Physics of the 
     Hungarian Academy of Sciences, H-1525 Budapest 114, Hungary$^{\ddag}$
\item[\mit] Massachusetts Institute of Technology, Cambridge, MA 02139, USA
\item[\debrecen] KLTE-ATOMKI, H-4010 Debrecen, Hungary$^\P$
\item[\florence] INFN Sezione di Firenze and University of Florence, 
     I-50125 Florence, Italy
\item[\cern] European Laboratory for Particle Physics, CERN, 
     CH-1211 Geneva 23, Switzerland
\item[\wl] World Laboratory, FBLJA  Project, CH-1211 Geneva 23, Switzerland
\item[\geneva] University of Geneva, CH-1211 Geneva 4, Switzerland
\item[\hefei] Chinese University of Science and Technology, USTC,
      Hefei, Anhui 230 029, China$^{\triangle}$
\item[\seft] SEFT, Research Institute for High Energy Physics, P.O. Box 9,
      SF-00014 Helsinki, Finland
\item[\lausanne] University of Lausanne, CH-1015 Lausanne, Switzerland
\item[\lecce] INFN-Sezione di Lecce and Universit\'a Degli Studi di Lecce,
     I-73100 Lecce, Italy
\item[\lyon] Institut de Physique Nucl\'eaire de Lyon, 
     IN2P3-CNRS,Universit\'e Claude Bernard, 
     F-69622 Villeurbanne, France
\item[\madrid] Centro de Investigaciones Energ{\'e}ticas, 
     Medioambientales y Tecnolog{\'\i}cas, CIEMAT, E-28040 Madrid,
     Spain${\flat}$ 
\item[\milan] INFN-Sezione di Milano, I-20133 Milan, Italy
\item[\moscow] Institute of Theoretical and Experimental Physics, ITEP, 
     Moscow, Russia
\item[\naples] INFN-Sezione di Napoli and University of Naples, 
     I-80125 Naples, Italy
\item[\cyprus] Department of Natural Sciences, University of Cyprus,
     Nicosia, Cyprus
\item[\nymegen] University of Nijmegen and NIKHEF, 
     NL-6525 ED Nijmegen, The Netherlands
\item[\caltech] California Institute of Technology, Pasadena, CA 91125, USA
\item[\perugia] INFN-Sezione di Perugia and Universit\'a Degli 
     Studi di Perugia, I-06100 Perugia, Italy   
\item[\cmu] Carnegie Mellon University, Pittsburgh, PA 15213, USA
\item[\prince] Princeton University, Princeton, NJ 08544, USA
\item[\rome] INFN-Sezione di Roma and University of Rome, ``La Sapienza",
     I-00185 Rome, Italy
\item[\peters] Nuclear Physics Institute, St. Petersburg, Russia
\item[\salerno] University and INFN, Salerno, I-84100 Salerno, Italy
\item[\ucsd] University of California, San Diego, CA 92093, USA
\item[\santiago] Dept. de Fisica de Particulas Elementales, Univ. de Santiago,
     E-15706 Santiago de Compostela, Spain
\item[\sofia] Bulgarian Academy of Sciences, Central Lab.~of 
     Mechatronics and Instrumentation, BU-1113 Sofia, Bulgaria
\item[\korea] Center for High Energy Physics, Adv.~Inst.~of Sciences
     and Technology, 305-701 Taejon,~Republic~of~{Korea}
\item[\alabama] University of Alabama, Tuscaloosa, AL 35486, USA
\item[\utrecht] Utrecht University and NIKHEF, NL-3584 CB Utrecht, 
     The Netherlands
\item[\purdue] Purdue University, West Lafayette, IN 47907, USA
\item[\psinst] Paul Scherrer Institut, PSI, CH-5232 Villigen, Switzerland
\item[\zeuthen] DESY, D-15738 Zeuthen, 
     FRG
\item[\eth] Eidgen\"ossische Technische Hochschule, ETH Z\"urich,
     CH-8093 Z\"urich, Switzerland
\item[\hamburg] University of Hamburg, D-22761 Hamburg, FRG
\item[\taiwan] National Central University, Chung-Li, Taiwan, China
\item[\tsinghua] Department of Physics, National Tsing Hua University,
      Taiwan, China
\item[\S]  Supported by the German Bundesministerium 
        f\"ur Bildung, Wissenschaft, Forschung und Technologie
\item[\ddag] Supported by the Hungarian OTKA fund under contract
numbers T019181, F023259 and T024011.
\item[\P] Also supported by the Hungarian OTKA fund under contract
  numbers T22238 and T026178.
\item[$\flat$] Supported also by the Comisi\'on Interministerial de Ciencia y 
        Tecnolog{\'\i}a.
\item[$\sharp$] Also supported by CONICET and Universidad Nacional de La Plata,
        CC 67, 1900 La Plata, Argentina.
\item[$\diamondsuit$] Also supported by Panjab University, Chandigarh-160014, 
        India.
\item[$\triangle$] Supported by the National Natural Science
  Foundation of China.
\item[\dag] Deceased.
\end{list}
}
\vfill






%
%

\begin{table}[th]
  \begin{center}
    \begin{tabular}{|c||r|r||r|r|}\hline
      Decay    &  \multicolumn{4}{|c|}{BRANCHING RATIOS} \\
      \cline{2-5}
      Channel  &  \multicolumn{2}{|c||}{$M=90\GeV$} &
                  \multicolumn{2}{|c|}{$M=180\GeV$}    \\
                 \cline{2-5}
               & $f=f'$ & $f=-f'$ & $f=f'$ & $f=-f'$  \\
      \hline
      $\rm \ell^* \rightarrow \ell \gamma$ &
                  89\%   &    0\%          &  37\%   &   0\%      \\
      \hline
      $\rm \ell^* \rightarrow \nu W      $ &
                  11\%   &   99\%          &  54\%   &  63\%      \\
      \hline
      $\rm \ell^* \rightarrow \ell Z     $ &
                   0\%   &    1\%          &   9\%   &  37\%      \\
      \hline
      \hline
      $\rm \nu^*  \rightarrow \nu  \gamma$ &
                   0\%   &   89\%          &   0\%   &  37\%      \\
      \hline
      $\rm \nu^*  \rightarrow \ell W     $ &
                  99\%   &   11\%          &  63\%   &  54\%      \\
      \hline
      $\rm \nu^*  \rightarrow \nu  Z     $ &
                   1\%   &    0\%          &  37\%   &   9\%      \\
      \hline
    \end{tabular}
    \icaption{Predicted branching ratios for charged and neutral
              excited lepton decays, for different choices of masses
              and couplings.
    \label{tab:bratio}}
  \end{center}
\end{table}

\begin{table}[th]
\begin{center}
\begin{tabular}{|c|c|c|c|c||c|c|c|c|}
\hline
  &  \multicolumn{4}{|c||}{Radiative Decays} & 
     \multicolumn{4}{c|}{Weak Decays}                 \\
\cline{2-9} 
& Signal & $N_D$ & $N_B$ & $\epsilon$  & Signal & $N_D$ & $N_B$ & $\epsilon$ \\ 

\hline \hline \rule{0pt}{13pt} 
& $\rm e^* e^* \rightarrow e e \gamma \gamma $                
&    0  &    0.6  &   49\%   
&  & &  &  \\

\cline{2-5} \rule{0pt}{13pt} 
& $\rm \mu^* \mu^* \rightarrow \mu \mu \gamma \gamma$ 
&    0     &     0.4  & 46\% 
& $\rm \ell^* \ell^* \rightarrow \nu \nu W W $
& 2710     &  2765    & 67\%          \\

\cline{2-5} \rule{0pt}{13pt} 
& $\rm \tau^* \tau^* \rightarrow \tau \tau \gamma \gamma$ 
&    1     &     0.2  & 40\% 
&  & & & \\

\cline{2-6} \cline{9-9} \rule{0pt}{13pt} 
& 
& & & & $\rm \nu_{\tau}^* \nu_{\tau}^* \rightarrow \tau \tau W W$
& & & 71\%   \\

\cline{6-9} \rule{0pt}{13pt} 
& $\rm \nu^* \nu^* \rightarrow \nu \nu \gamma \gamma$
&    2     &     1.6  & 44\% & $\rm \nu_{e}^* \nu_{e}^* \rightarrow e e W W$
&    1     &     0.12 & 18\%          \\

\cline{6-9} \rule{0pt}{13pt} \raisebox{2.0mm}[0mm][0mm]
{\rotatebox{90} {Pair Production}} 
& 
& & & & $\rm \nu_{\mu}^* \nu_{\mu}^* \rightarrow \mu \mu W W$
&    1     &     0.34 & 21\%          \\

\hline \hline \rule{0pt}{13pt} 
& $\rm e^* e \rightarrow e e \gamma$      &  563     &   564    & 63\% 
& $\rm e^* e \rightarrow \nu_e W e$ &  452  &   455    & 24\%     \\

\cline{2-9} \rule{0pt}{13pt} 
& $\rm \mu^* \mu \rightarrow \mu \mu \gamma$ & 71 & 64 & 61\% 
& $\rm \mu^* \mu \rightarrow \nu_{\mu} W \mu$
&  476     &   479    & 49\%          \\

\cline{2-9} \rule{0pt}{13pt} 
& $\rm \tau^* \tau \rightarrow \tau \tau \gamma$ 
&   64     &    55    & 43\% 
& $\rm \tau^* \tau \rightarrow \nu_{\tau} W \tau$
& 1004     &   972    & 44\%          \\

\cline{2-9} \rule{0pt}{13pt} 
& 
&          &          &      
& $\rm \nu_{e}^* \nu_{e} \rightarrow e W \nu_{e}$
&  452     &   455    & 47\%          \\

\cline{6-9} \rule{0pt}{13pt} 
& $\rm \nu^* \nu \rightarrow \nu \nu \gamma$ 
&  191     &   219    & 66\% 
& $\rm \nu_{\mu}^* \nu_{\mu} \rightarrow \mu W \nu_{\mu}$
&  476     &   479    & 51\%          \\

\cline{6-9} \rule{0pt}{13pt} \raisebox{0.0mm}[0mm][0mm]
{\rotatebox{90} {Single Production}} 
& 
&          &          &      
& $\rm \nu_{\tau}^* \nu_{\tau} \rightarrow \tau W \nu_{\tau}$
& 1004     &   972    & 41\%          \\

\hline
\end{tabular}
\icaption{Number of candidates $N_D$, number of background events $N_B$, 
          and average signal efficiencies $\epsilon$, 
          for radiative and weak decays, in the pair
          production (upper part) and the single production
          (lower part) searches.
\label{tab:selections}}
\end{center}
\end{table}
\newpage

\begin{table}[th]
\begin{center}
\begin{tabular}{|c||c|c|c|}
\hline
    Excited           &  \multicolumn{3}{|c||}{95\% CL Mass Limit (\GeV)} \\
               \cline{2-4}
    Lepton   & $f=f'$ & $f=-f'$ 
             & Coupling Independent \\
    \hline
~~~~$\rm \e^*~~~~    $  &  94.2  &  92.6  & 92.4 \\
    \hline
    $\rm \mu^*        $ &  94.2  &  92.6  & 92.4 \\
    \hline
    $\rm \tau^*       $ &  94.2  &  92.6  & 91.7 \\
    \hline
    \hline
    $\rm \nu^*_e      $ &  93.9  &  94.1  & 93.4 \\
    \hline
    $\rm \nu^*_{\mu}  $ &  94.0  &  94.1  & 93.5 \\
    \hline
    $\rm \nu^*_{\tau} $ &  91.5  &  94.1  & 90.2 \\
    \hline
\end{tabular}
\icaption{95\% confidence level lower mass limits for the different
          excited leptons obtained from pair production searches. For each
          flavour, the mass limits for $f=f'$, $f=-f'$ and for the
          coupling independent case, are shown.
\label{tab:limits}}
\end{center}
\end{table}

\clearpage
%
%

\clearpage

\clearpage

\clearpage


\begin{thebibliography}{99}

\bibitem{yellow}
F. Boudjema {\it et al.}, ``Z Physics at LEP 1'', Vol.2, ed. J. Ellis 
{\it et al.}, CERN 89-08 (1989) 188 and references therein.

\bibitem{LEP_HERA}
ALEPH Coll.,  D. Buskulic {\it et al.}, Phys. Lett.   {\bf B385} (1996) 445;\\
DELPHI Coll., P. Abreu {\it et al.},    E. Phys. J.   {\bf C8}   (1999)  41;\\
L3 Coll.,     M. Acciarri {\it et al.}, Phys. Lett.   {\bf B401} (1997) 139;\\
OPAL Coll.,   K. Ackerstaff {\it et al.}, Eur. Phys. J. {\bf C1} (1998)  45;\\
H1 Coll.,     S. Aid {\it et al.},      Nucl. Phys.   {\bf B483} (1997)  44;\\
ZEUS Coll.,   J. Breitweg {\it et al.}, Z. Phys.      {\bf C76}  (1997) 631.

\bibitem{hagi}    
K. Hagiwara {\it et al.}, Z. Phys. {\bf C29} (1985) 115.        

\bibitem{neus}
F. Boudjema  {\it et al.}, Z. Phys. {\bf C57} (1993) 425; \\
F. Boudjema  {\it et al.}, Phys. Lett. {\bf B240} (1990) 485; \\
M.B. Voloshin, {\it et al.}, Sov. Phys. JETP {\bf 64} (1986) 446; \\
M.B. Voloshin, Phys. Lett. {\bf B209} (1988) 360.        

\bibitem{l3-detector}   
L3 Coll., B. Adeva {\it et al.}, 
Nucl. Instr. and Meth. {\bf A289} (1990) 35; \\
L3 Coll., M. Chemarin {\it et al.}, 
Nucl. Instr. and Meth. {\bf A349} (1994) 345; \\
L3 Coll., M. Acciarri {\it et al.}, 
Nucl. Instr. and Meth. {\bf A351} (1994) 300; \\
L3 Coll., A. Adam {\it et al.}, 
Nucl. Instr. and Meth. {\bf A383} (1996) 342.

\bibitem{bhwide}
S. Jadach, {\it et al.}, Phys. Lett. {\bf B390} (1997) 298.

\bibitem{tee} 
D. Karlen, Nucl. Phys. {\bf B289} (1987) 23.

\bibitem{koralz}
S. Jadach {\it et al.}, Comp. Phys. Comm. {\bf 79} (1994) 503.

\bibitem{ggg}  
F.A. Berends {\it et al.}, Nucl. Phys. {\bf B186} (1981) 22; \\
CALKUL Coll., F.A. Berends {\it et al.}, Nucl. Phys. {\bf B239} (1984) 395.

\bibitem{koralw} KORALW: Version 1.21 is used.
  M. Skrzypek, {\it et al.}, Comp. Phys. Comm. {\bf 94} (1996) 216; \\
  M. Skrzypek, {\it et al.}, Phys. Lett. {\bf B372} (1996) 289.

\bibitem{pythia} PYTHIA~5.7 and JETSET~7.4 Physics and Manual, 
        T. Sj{\"o}strand, CERN-TH/7112/93 (1993), revised August 1995; 
        Comp. Phys. Comm. {\bf 82} (1994) 74.

\bibitem{exca} 
F.A. Berends, {\it et al.}, Nucl. Phys. {\bf B424} (1994) 308; \\
F.A. Berends, {\it et al.}, Nucl. Phys. {\bf B426} (1994) 344; \\
F.A. Berends, {\it et al.}, Nucl. Phys. (Proc. Suppl.) {\bf B37} (1994) 163; \\
R. Kleiss {\it et al.}, Comp. Phys. Comm. {\bf 85} (1995) 447; \\
R. Pittau, Phys. Lett. {\bf B335} (1994) 490.

\bibitem{geant} The L3 detector simulation is based on GEANT Version 3.15,
R. Brun {\it et al.}, CERN DD/EE/84-1 (Revised 1987); \\
The GHEISHA program (H. Fesefeldt, RWTH Aachen Report PITHA 85/02 (1985))
is used to simulate hadronic interactions.

\end{thebibliography}
\end{document}